\documentclass{aa}
\usepackage{natbib}
\usepackage{graphicx}
\usepackage[varg]{txfonts}

\bibpunct{(}{)}{;}{a}{}{,} 

\begin{document}

\title{Induced gravitational collapse at extreme cosmological distances: the case of GRB 090423}
\titlerunning{IGC at high redshift}
\authorrunning{Ruffini et al.}
\author{R. Ruffini$^{1,2,3,4}$, L. Izzo$^{1,2}$,  M. Muccino$^{1}$,   G. B. Pisani$^{1,3}$ , J. A. Rueda$^{1,2,4}$, Y. Wang$^{1}$, C. Barbarino$^{1,5}$, C. L. Bianco$^{1,2}$, M. Enderli$^{1,3}$, M. Kovacevic$^{1,3}$}
\offprints{ruffini@icra.it}
\institute{$^{1}$Dip. di Fisica and ICRA, Sapienza Universit\`a di Roma, Piazzale Aldo Moro 5, I-00185 Rome, Italy.\\
$^{2}$ICRANet, Piazza della Repubblica 10, I-65122 Pescara, Italy.\\
$^{3}$Universit\'e de Nice Sophia Antipolis, CEDEX 2, Grand Ch\^{a}teau Parc Valrose, Nice, France.\\
$^{4}$ICRANet-Rio, Centro Brasileiro de Pesquisas Fisicas, Rua Dr. Xavier Sigaud 150, Rio de Janeiro, RJ, 22290-180, Brazil.\\
$^{5}$INAF-Napoli, Osservatorio Astronomico di Capodimonte, Salita Moiariello 16, I-80131 Napoli, Italy.\\ }
\date{}

\abstract
{The induced gravitational collapse (IGC) scenario has been introduced in order to explain the most energetic gamma ray bursts (GRBs),  $E_{iso}  =  10^{52} - 10^{54}$ erg, associated with type Ib/c supernovae (SNe). It has led to the concept of binary-driven hypernovae (BdHNe) originating in a tight binary system composed by a FeCO core on the verge of a SN explosion and a companion neutron star (NS). Their evolution is characterized by a rapid sequence of events: 1) The SN explodes, giving birth to a new NS ($\nu$NS). The accretion of SN ejecta onto the companion NS increases its mass up to the critical value; 2) The consequent gravitational collapse is triggered, leading to the formation of a black hole (BH) with GRB emission; 3) A novel feature responsible for the emission in the GeV, X-ray, and optical energy range occurs and is characterized by specific power-law behavior in their luminosity evolution and total spectrum; 4) The optical observations of the SN then occurs.}
{We investigate whether GRB 090423, one of the farthest observed GRB at $z=8.2$, is a member of the BdHN family.}   
{We compare and contrast the spectra, the luminosity evolution, and the detectability in the observations by \textit{Swift} of GRB 090423 with the corresponding ones of the best known BdHN case, GRB 090618.}
{Identification of constant slope power-law behavior in the late X-ray emission of GRB 090423 and its overlapping with the corresponding one in GRB 090618, measured in a common rest frame, represents the main result of this article. This result represents a very significant step on the way to using the scaling law properties, proven in Episode 3 of this BdHN family, as a cosmological standard candle.}
{Having identified GRB 090423 as a member of the BdHN family, we can conclude that SN events, leading to NS formation, can already occur already at $z=8.2$, namely at 650 Myr after the Big Bang. It is then possible that these BdHNe originate stem from 40-60 M$_{\odot}$ binaries.  They are probing the Population II stars after the completion and possible disappearance of Population III stars.}

\maketitle

\section{Introduction}

The induced gravitational collapse (IGC) paradigm \citep{Ruffini2011,Rueda2012,Izzo2012} has been proposed to explain a class of very energetic ($E_{iso} \sim 10^{52}$--$10^{54}$ erg) long gamma ray bursts (GRBs) associated with supernovae (SNe). A new class of systems, with progenitor a tight binary composed by a FeCO core and a companion neutron star (NS), has been considered.
These systems evolve in a very rapid sequence lasting a few hundred seconds in their rest frame: 1) the SN explodes giving birth to a new NS ($\nu$NS); 2) the accretion of the SN ejecta onto the companion NS increases its mass, reaching the critical value; 3) the gravitational collapse is triggered, leading to the formation of a black hole (BH) with GRB emission. Such systems have been called binary-driven hypernovae \citep[BdHN][]{RuffiniMuccino2014}.

Observationally, this authentic cosmic matrix is characterized by four distinct episodes, with the ``in'' state represented by a FeCO core and a NS and the ``out'' state by a $\nu$NS and a BH. Each episode contains specific signatures in its spectrum and luminosity evolution. Up to now, the IGC paradigm has been verified in a dozen GRBs, all with redshift up to $z \sim 1$ \citep{Izzo2012b,Penacchioni2012,Penacchioni2013,Pisani2013,Ruffini2013}.

Various approaches have been followed to reach an understanding of long GRBs. One of these has been the use of statistical tools to obtain general results that examine the most complete source catalog \citep[see, e.g.,][and references therein]{Nousek2006,Kann2011,Salvaterra2012,Margutti2013}.

We follow a different approach here. We first identified the specific class of BdHNe of GRBs related to SNe, as mentioned above, widely tested at $z \approx 1$. We furthermore explore the members of this class by extending our analysis to higher values of the cosmological redshifts. We do that by taking the scaling laws for the cosmological transformations into account, as well as the specific sensitivities of the GRB detectors (in this case \textit{Swift}, \citealp{Gehrels2005}, and \textit{Fermi}, \citealp{Meegan2009}).

Our aim is to verify that  such BdHNe, originating in a SN and a companion NS, did form in the earliest phases of the universe. If this is confirmed, we go on to examine the possibility that all GRBs with $E_{iso} \sim 10^{52}-10^{54}$ erg are indeed associated to SN and belong to the BdHN family independently of their space and time location. 

\section{The four episodes of BdHNe sources}

In order to achieve this goal, we recall the four above-mentioned episodes, present in the most general BdHN (see Fig. \ref{fig:Maxime}): 

\begin{figure}
\centering
\includegraphics[width=0.99\hsize,clip]{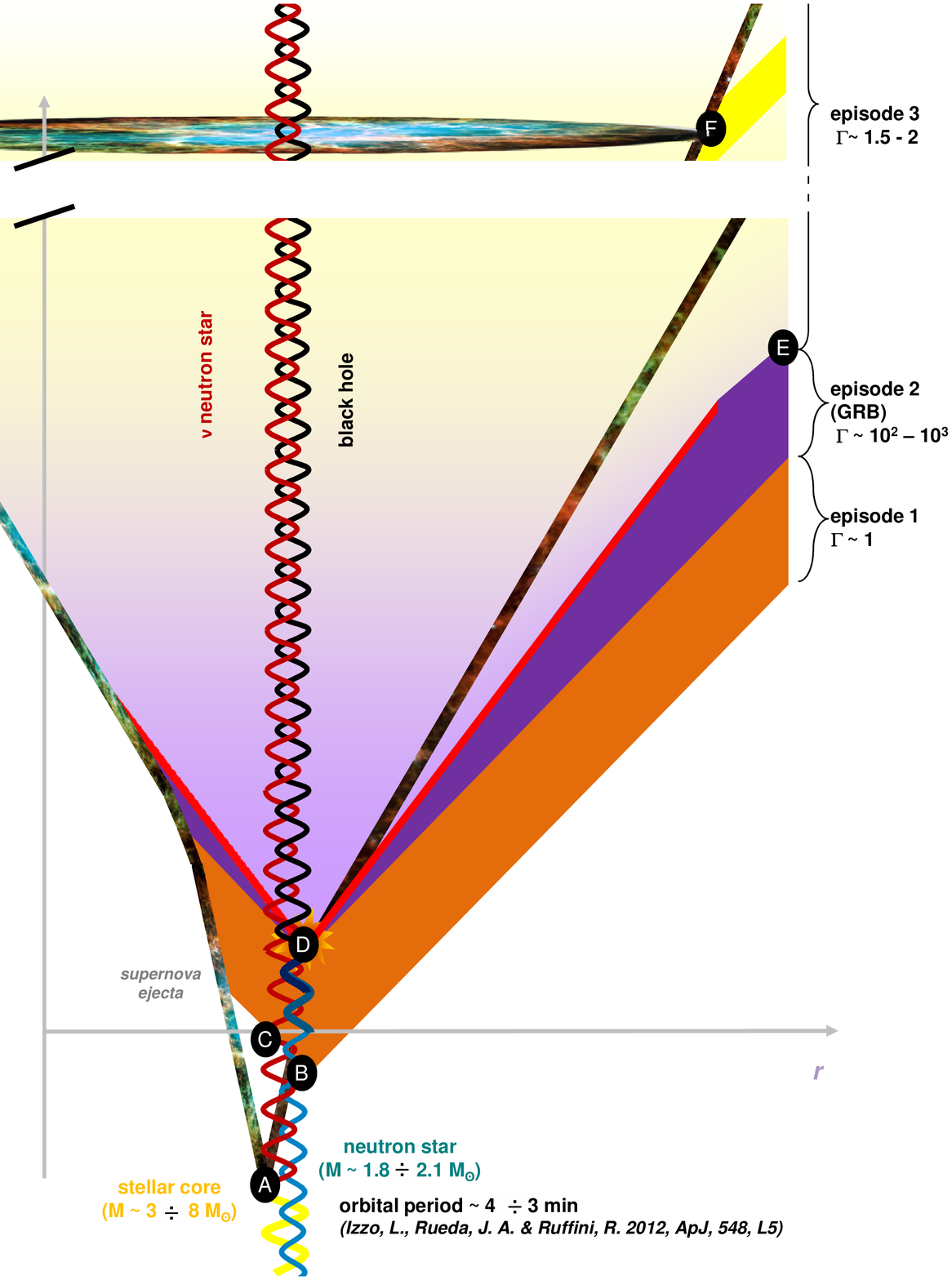}
\caption{Space-time diagram of the induced gravitational collapse applied to GRB 090618 \citep{EnderliTEXAS2013,RuffiniTEXAS2013}. The sequence is summarized as follows: A) the explosion as a SN of the evolved FeCO core which creates a $\nu$-NS and its remnant; B) the beginning of the accretion of the SN ejecta onto the companion NS, emitting Episode 1; C) a prolonged interaction between the $\nu$-NS and the NS binary companion; D) the companion NS reaches its critical mass by accretion, and a BH is formed with the consequent emission of a GRB; E) the arrival time at the separatrix between Episodes 2 and 3; F) the optical emission of the SN due to the decay of $^{56}$Ni after $t_a^d\sim10(1+z)$ days in the observer frame (Episode 4).}
\label{fig:Maxime}
\end{figure}

\begin{figure}
\centering
\includegraphics[width=0.99\hsize,clip]{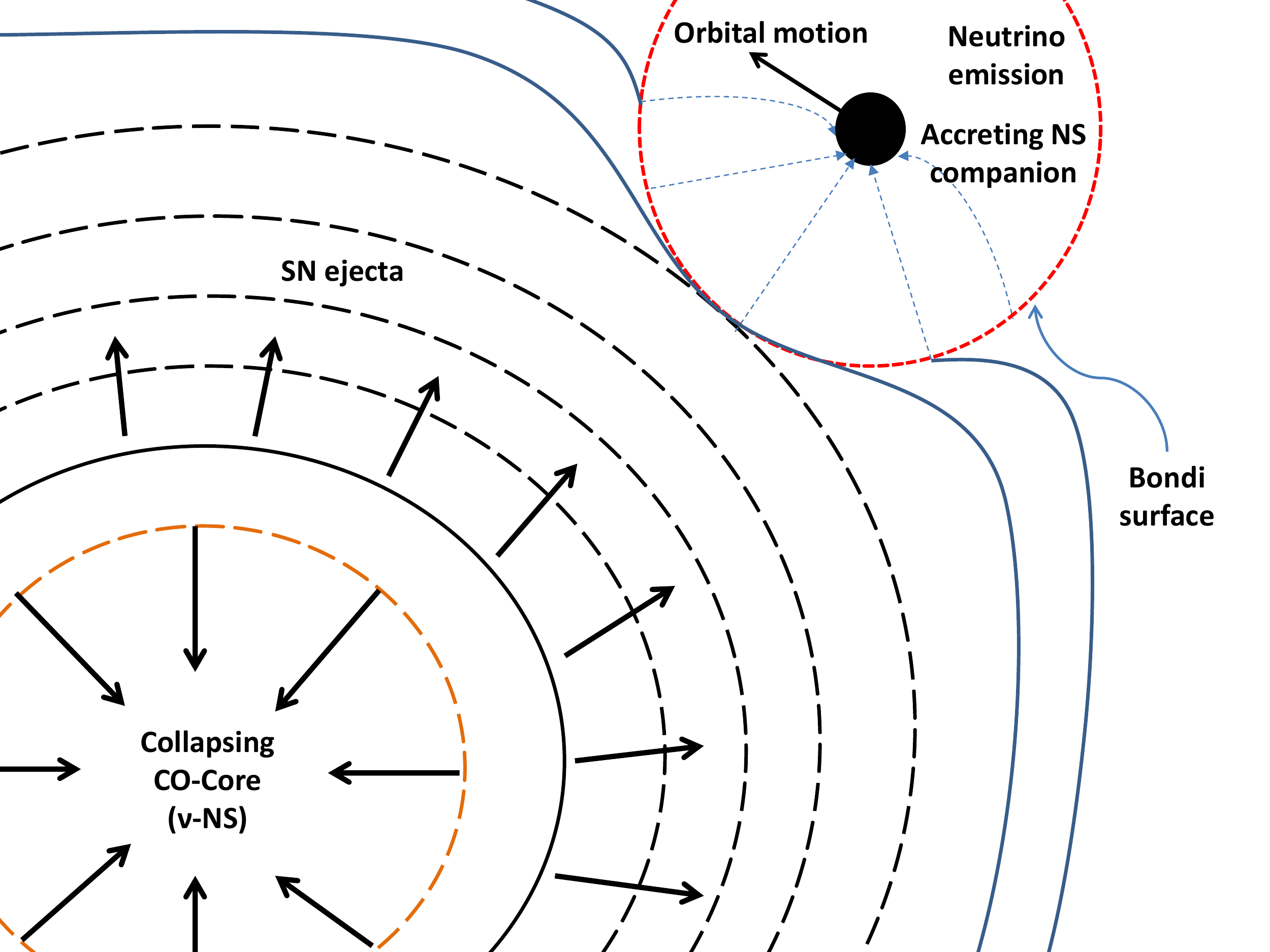}
\caption{A sketch (not in scale) of the accretion induced gravitational collapse (IGC) scenario.}
\label{fig:Jorge}
\end{figure}

\textit{Episode 1} has the imprint of the onset of a SN in the tight binary system with the companion neutron star (NS) (see Fig. \ref{fig:Jorge}). It stemmed from the hyper-critical accretion of the SN matter ejecta ($\sim 10^{-2} M_\odot~s^{-1}$) \citep{Rueda2012}. Decades of conceptual progress have passed from the original work of \citet{BondiHoyle} and \citet{Bondi1952} to the problem of a ``hypercritical'' accretion rate. 
This problem has acquired growing scientific interest as it moved from the classical astronomical field to the domain of the relativistic astrophysics. 
The crucial role of neutrino cooling, earlier considered by \citet{Zeldovich1972} and later on by \citet{Bisnovati1984} in SN fallback, has been recognized to play a crucial role in describing binary common envelope systems by \citet{Chevalier1989,Chevalier1993}. 
In the work by \citet{Fryer1996}, and more recently in \citet{Fryer2009}, it was clearly shown that an accretion rate $\dot{M} \sim 10^{-2} M_\odot$ s$^{-1}$ onto a neutron star (NS) could lead in a few seconds to the formation of a black hole (BH), when neutrino physics in the description of the accreting NS is taken into due account. The data acquired in Episode 1 of GRB 090618 \citep{Izzo2012b}, as well as the one in GRB 101023 \citep{Penacchioni2012}, GRB 110709B \citep{Penacchioni2013}, and GRB 970828 \citep{Ruffini2013}, give for the first time the possibility to probe the Bondi-Hoyle hypercritical accretion and possibly the associated neutrino emission, which was theoretically considered by \citet{Zeldovich1972,Chevalier1993,Fryer1996}, and \citet{Fryer2009}.

\textit{Episode 2} is the canonical GRB emission, which originated in the collapse of the companion NS, which reached its critical mass by accretion of the SN ejecta and then collapsed to a black hole (BH), indeed emitting the GRB. 

\begin{figure*}
\centering
\includegraphics[width=0.99\hsize,clip]{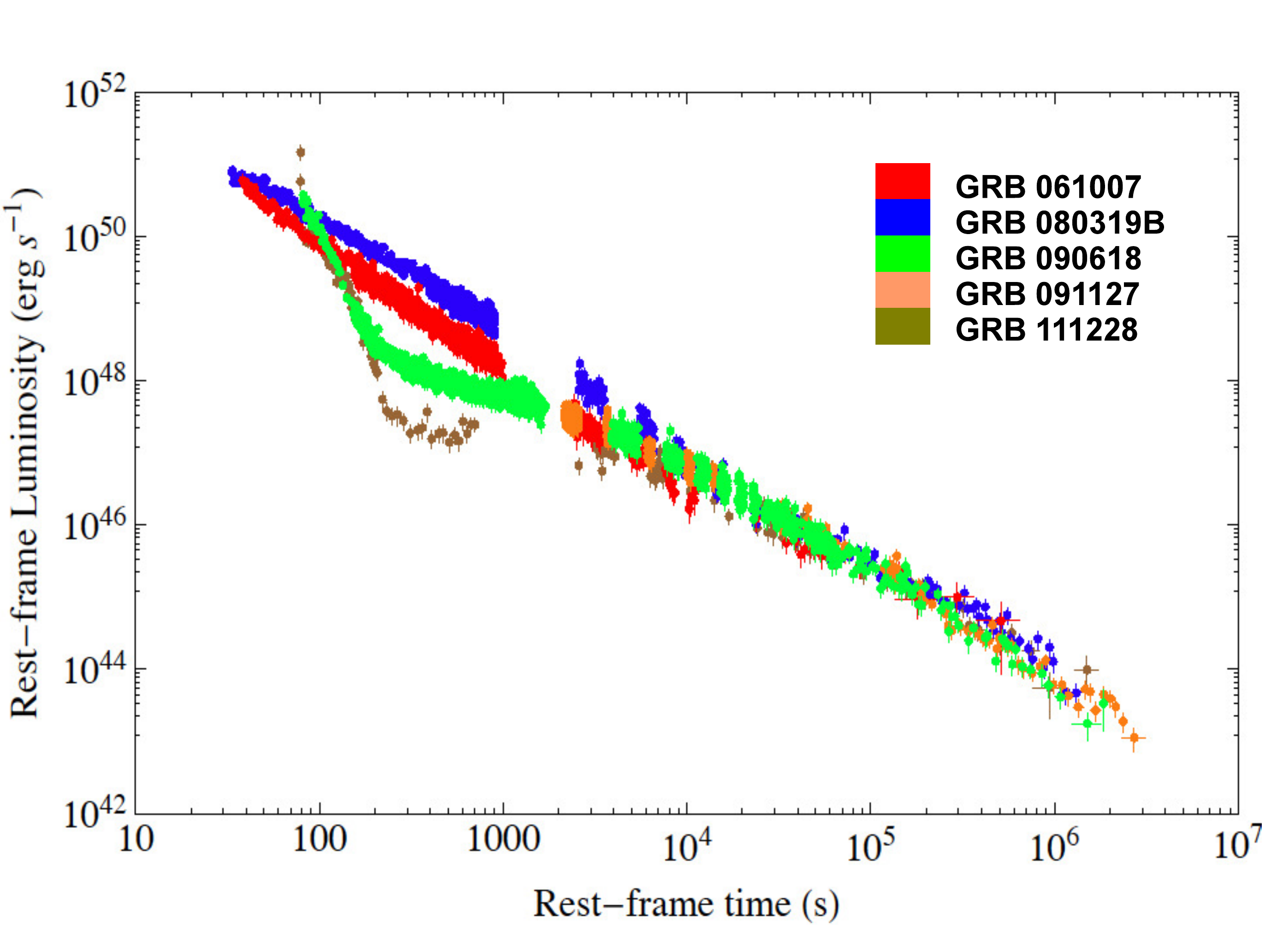}
\caption{Rest-frame, X-ray afterglow, luminosity light curves of some IGC GRBs-SNe belonging to the ``golden sample'' described in \citet{Pisani2013}. The overlapping after $10^4$ s is clearly evident, confirming the presence of an Episode 3 in this GRB.}
\label{fig:no0}
\end{figure*}

\begin{figure*}
\centering
\includegraphics[width=0.47\hsize,clip]{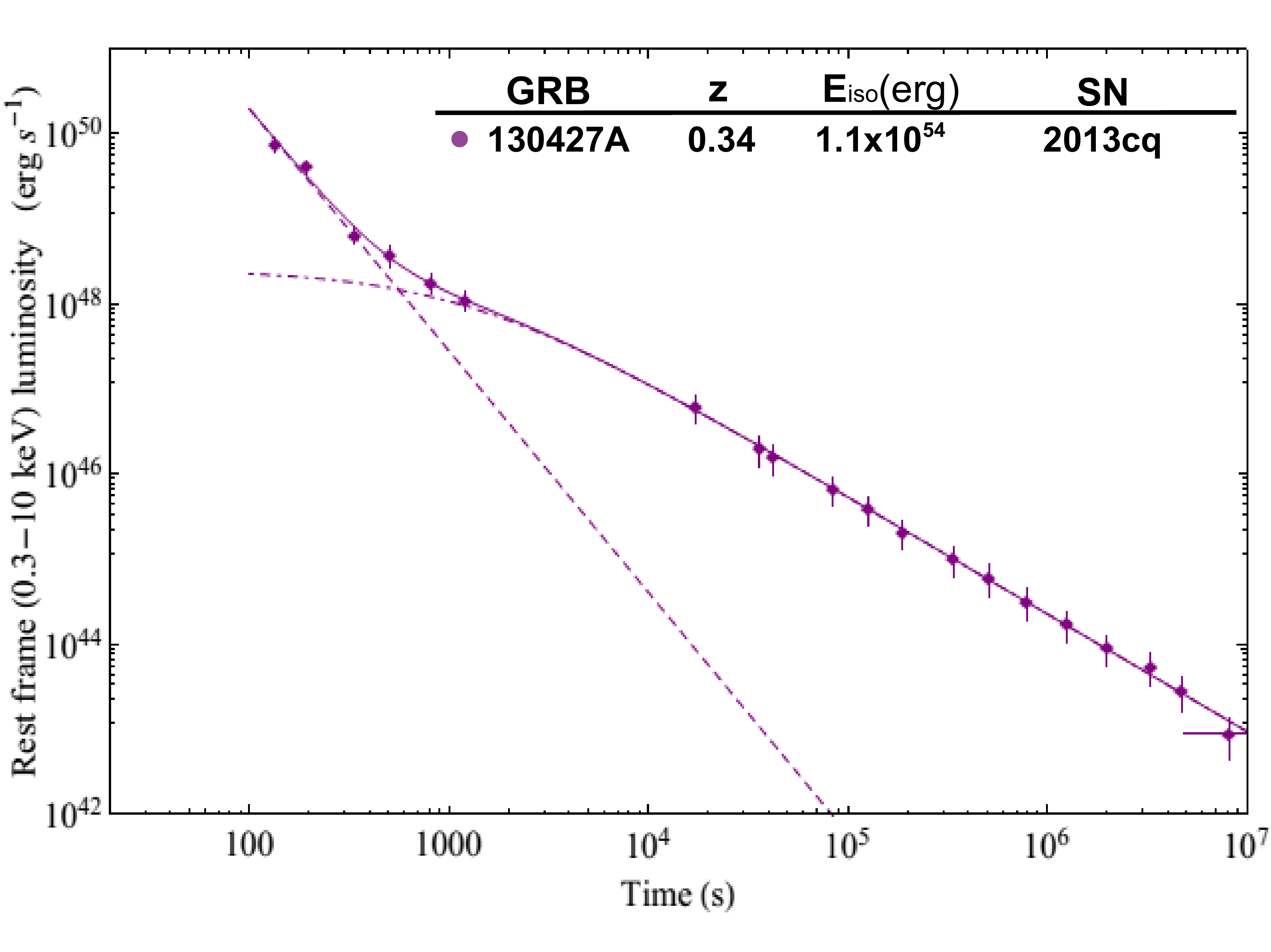}
\includegraphics[width=0.47\hsize,clip]{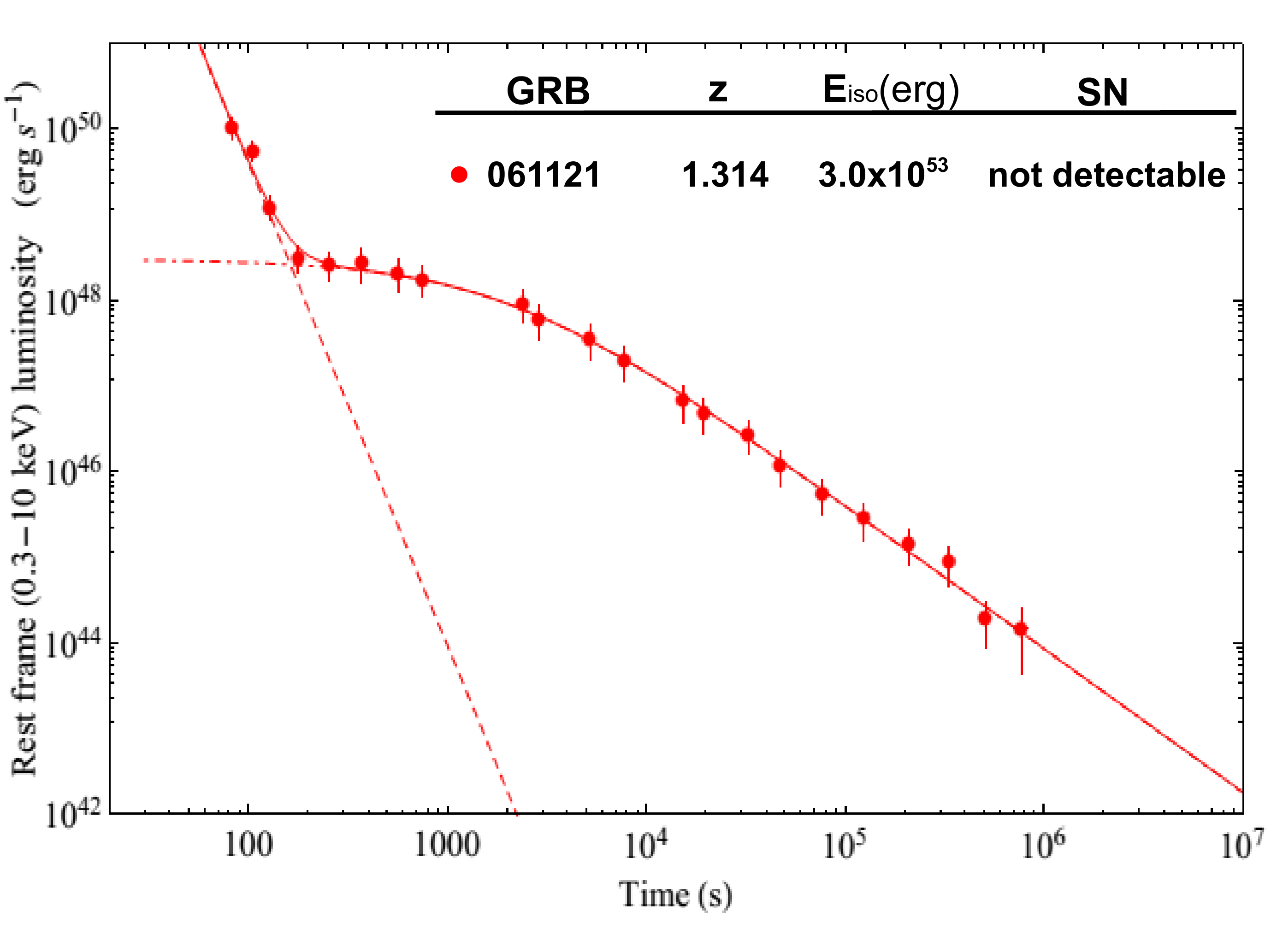}\\
\includegraphics[width=0.47\hsize,clip]{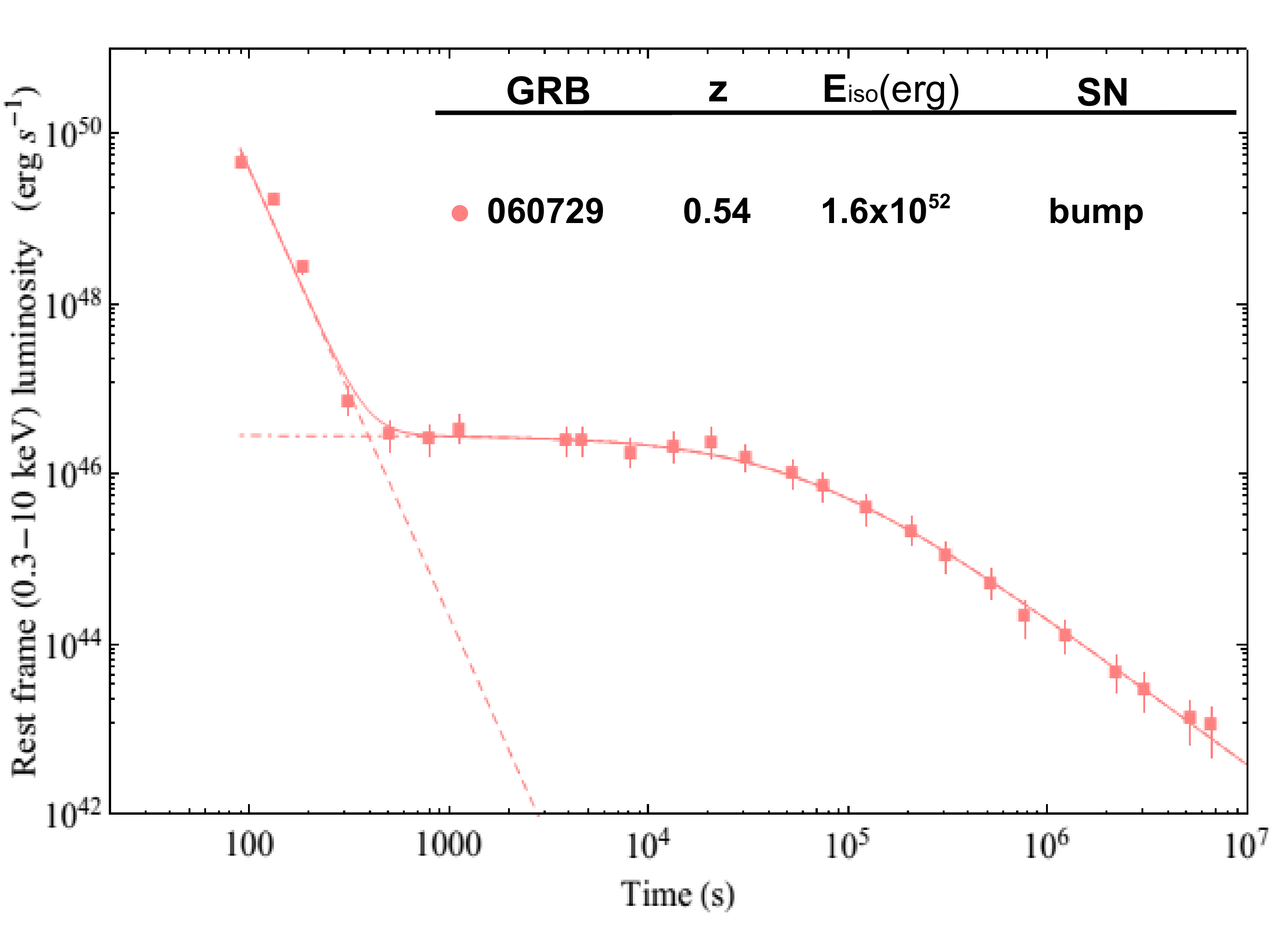}
\includegraphics[width=0.47\hsize,clip]{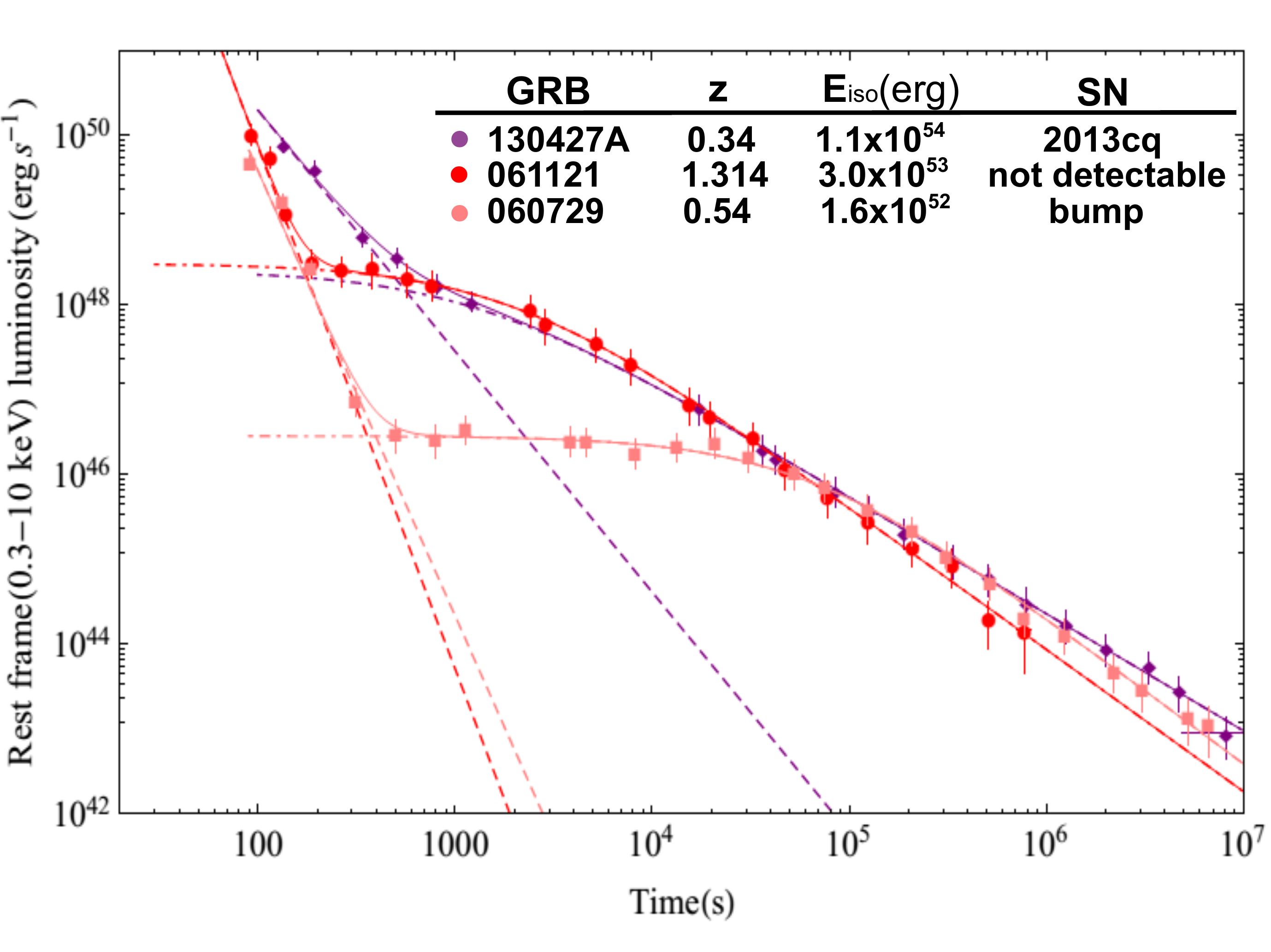}\\
\caption{Rest-frame, ($0.3$ -- $10$) keV, and re-binned luminosity light curves of GRB 130427A (upper left), GRB 061121 (upper right), GRB 060729 (lower left) and a combined picture (lower right). The fits to their emission is done using a power-law function for the early steep decay and a phenomenological function for the following emission, which is described well in \citet{RuffiniMuccino2014}.}
\label{fig:nest}
\end{figure*}

\textit{Episode 3}, observed in X-rays by Swift-XRT, shows very precise behavior consisting of steep decay, starting at the end point of the prompt emission, and then a plateau phase followed by a late power-law decay (see \citealp{Pisani2013} and also Fig. \ref{fig:no0}). The late X-ray luminosities of BdHNe, in their rest-frame energy band $0.3$--$10$ keV, show a common power-law behavior with a constant decay index clustering around $\alpha = -1.5 \pm 0.2$.
The occurrence of such a constant afterglow decay has been observed in all the BdHN sources examined. For example, see in Fig.~\ref{fig:nest} the data for GRB 130427A, GRB 061121, GRB 060729, respectively.
It appears an authentic nested structure, in the late X-ray emission of GRBs associated to SNe, and it has indeed to be indicated as the qualifying feature for a GRB to be a member of the BdHNe family \citep{RuffiniMuccino2014}. It is clear that such a phenomenon offers a strong challenge for explaining by any GRB model. 

In addition to these X-ray features, the observations of GRB 130427A by the \textit{Swift}, \textit{Fermi}, and \textit{Konus-WIND} satellites and a large number of optical telescopes have led to the evidence of such power laws at very high energies, in $\gamma$-rays and at the optical wavelengths \citep[][see also \citet{RuffiniWang2014}]{Fermi130427A,Melandri2014}.

\textit{Episode 4} is characterized by the emergence of the SN emission after about $10$--$15$ days from the occurrence of the GRB in the rest frame of the source, which has been observed for almost all the sources fulfilling the IGC paradigm with $z \sim 1$.

\section{GRB 090423 compared and contrasted with GRB 090618}

We first consider the data of GRB 090423, one of the farthest GRB ever observed at $z=8.2$ \citep{Salvaterra2009,Tanvir2009}, with the prototypical member of the BdHNe class, namely GRB 090618, and its associated SN \citep{Izzo2012b}. In other words we proceed with a specific ansatz: we verify that GRB 090423, at $z=8.2$, presents analogous intrinsic features to GRB 090618, which was observed at $z=0.54$. 

We proceed by examining (see Section 4) each one of the above episodes for both sources,  by a detailed spectral analysis and simulations. We first verify that Episode 1 of GRB 090618 transposed at redshift $z=8.2$ should not have triggered the Swift BAT detector. Indeed, no precursor in the light curve of GRB 090423 was detected. Consequently, we do not address any theoretical considerations of the hypercritical accretion in Episode 1 of GRB 090423, since it is not observable in this source (see Section 5).
We also notice that the distance of GRB 090423 prevents any possible detection of a SN associated with this GRB, and therefore Episode 4 cannot be observed in GRB 090423.

For Episode 2, we have found that indeed the transposed emission of GRB 090618 should provide a positive trigger: we show in Section 6 that the duration, the observed luminosity and the spectral emission of Episode 2 in GRB 090423 present analogous intrinsic features to the transposed ones of GRB 090618 and differ only in the spectral energy distribution due to different circumburst medium properties.

For Episode 3, the crucial result, probing the validity of the above ansatz, is that the late X-ray emission in GRB 090423, computed in the rest frame of the burst at $z=8.2$, precisely coincides (overlaps) with the corresponding late X-ray emission in GRB 090618, as evaluated in the rest frame of the source at $z=0.54$, see Section 7. The occurrence of this extraordinary coincidence in Episode 3 proves that GRB 090423 is indeed a member of the BdHN family. This in particular opens the possibility of elaborating a role for the late X-ray emission in BdHNe as a standard candle.

\section{The data}

GRB 090423 was discovered on 23 April 2009, 07:55:19 UT, $T_0$ from here, by the \emph{Swift} Burst Alert Telescope (BAT)  \citep{Krimm}, at coordinates R.A. = 09$^h$ 55$^m$ 35$^s$, Dec = +18$^{\circ}$  09$'$ 37$''$ (J2000.0; 3$'$ at $90$ \% containment radius). The \emph{Swift}-BAT light curve showed a double-peaked structure with a duration of about 20 s. The X-ray Telescope (XRT) \citep{Burrows2005} on board the same spacecraft started to observe GRB 090423 72.5 s after the initial trigger, finding a fading source and providing enhanced coordinates for the follow-up by on-ground telescopes that have allowed the discovery of its redshift \citep[$z = 8.2$,][]{Salvaterra2009,Tanvir2009}. The light curve is characterized by an intense and long flare peaking at about $T_0 + 180$, followed by a power-law decay, observed from the second orbit of Swift \citep{Stratta2009}. 
The prompt emission from GRB 090423 was also detected by the Fermi Gamma-Ray Burst Monitor (GBM, trigger 262166127 / 090423330), \citep{vonKienlin2009}, whose on-ground location was consistent with the \emph{Swift} position.
The Large Area Telescope (LAT) on-board the Fermi satellite did not detected any signal from GRB 090423. The GBM light curve showed a single-structured peak with a duration of about 12 s, whose spectral energy distribution was best fit with a power law with an exponential cut-off energy, parameterized as $E_{peak}=(82 \pm 15)$ keV. The observed fluence was computed from  Fermi data to be $S_{\gamma} = 1.1 \times 10^{-6} \, $ ergs/cm$^2$ that, considering the standard $\Lambda$CDM cosmological model, corresponds to an isotropic energy emitted of $E_{iso} = 1.1 \times 10^{53}$ ergs for the spectroscopic redshift $z = 8.2$ \citep{vonKienlin2009b}.
With these values for $E_{peak}$ and $E_{iso}$, GRB 090423 satisfies the Amati relation, which is only valid for long GRBs \citep{Amati2002}.

\section{The impossibility of detecting Episode 1}

It has become natural to ask if observations of Episodes 1 and 2 in the hard X-ray energy range could be addressed for the case of GRB 090423. We have first analyzed a possible signature of Episode 1 in GRB 090423. Since the Swift-BAT, (15 - 150) keV, light curve is a single-structured peak with duration of $\sim$ 19 s, as detected by Swift BAT, with no thermal emission in its spectrum and no detection of any emission from a precursor in the Swift and Fermi data, we have considered the definite possibility that Episode 1 was not observed at all. In this light, the best way to check this possibility consists in verifying that the Episode 1 emission is below the threshold of the Swift BAT detector, consequently, it could have not triggered the Swift BAT. We have considered the prototype of Episode 1 as the one observed in GRB 090618 \citep{Izzo2012}, which is at redshift $z$ = 0.54, and then we transposed it at redshift $z=8.2$, simulating the observed emission of GRB 090618 as if it had been observed at this large distance. Then, we performed a time-resolved spectral analysis of Episode 1 in GRB 090618, using a Band function as spectral model, and finally we translated the specific photon spectra obtained from the analysis at the redshift of GRB 090423. This last operation consists in two transformations, concerning the peak energy $E_{peak}$ of the Band function and the normalization value $K_{Band}$. The new value of the peak energy is simply given by $E_{peak,8}$ = $E_{peak}$ $(1+0.54)/(1+8.2)$, while the normalization, which corresponds to the specific photon flux at 1keV, requires knowledge of the luminosity distances of the two bursts, $d_l(z)$ :

\begin{equation}
K_{Band,8} = K_{Band} \left(\frac{1+8.2}{1+0.54}\right)^2 \left(\frac{d_l (0.54)}{d_l (8.2)}\right)^2.
\end{equation} 

Another transformation concerns the observational time of Episode 1 of GRB 090618 at redshift $z$ = 8.2. At large distances, any astrophysical event will be dilated in time by the cosmological redshift effect, which in the current case modifies the time interval by a quantity $(1+8.2)/(1+0.54)$ = $5.97$. The knowledge of this time interval is fundamental since it represents the exposure of a simulated spectrum translated at $z$ = 8.2. We considered Fermi GBM data for analyzing the time-resolved spectra of GRB 090618, as described by \citet{Izzo2012}. The wide energy range of Fermi GBM NaI detectors, (8 - 1000) keV, allows a more accurate determination of the Band parameters, which are used as input values for the simulated spectra. We also rebinned the Fermi data considering a signal-to-noise ratio (SNR) = 10, and finally performed our spectral analysis. The next step consisted in transforming the peak energy of the Band function and of the normalization of all these time-resolved photon spectra $N(E)$, as described above. 

Following the work of \citet{Band2003}, the sensitivity of an instrument to detect a burst depends on its burst trigger algorithm. The Swift BAT trigger algorithm, in particular, looks for excesses in the detector count rate above expected background and constant sources. There are several criteria for determining the correct BAT threshold significance $\sigma_0$ for a single GRB \citep{Barthelmy2005}, but in this work we have considered the treatment given in \citet{Band2003}. Recently, the threshold of Swift BAT has been modified to allow detecting of subthreshold events, but since GRB 090423 was detected before, the \citet{Band2003} analysis is still valid for our purposes. The preset threshold significance for Swift BAT can be expressed by the following formula:

\begin{equation}\label{eq:no2}
\sigma_0 = \frac{A_{eff} f_{det} f_{mask} \Delta t \int_{15}^{150} \epsilon(E) N(E) d E}{\sqrt{A_{eff} f_{det} \Delta t \int_{15}^{150} B(E) d E}},
\end{equation}
where $A_{eff}$ is the effective area of the detector, $f_{det}$ the fraction of the detector plane that is active, $f_{mask}$ the fraction of the coded mask that is open, $\Delta t$ the exposure of the photon spectrum $N(E)$, $\epsilon (E)$ the efficiency of the detector, and $B(E)$ the background. We considered the values for these parameters as the ones given in the Band work (with the exception of the detecting area, assumed to be $A_{eff} = 5200 cm^2$), while the efficiency and the background were obtained from the Swift BAT integrated spectrum of GRB 090423 using the XSPEC fitting package. Then we considered as input photon spectra $N(E)$ the ones obtained from the Fermi GBM analysis of Episode 1 of GRB 090618 and translated for the redshift $z$ = 8.2.
It is appropriate to note that the transformations of spectra presented above are the correct ones: since the sensitivity of Swift-BAT strongly depends on the peak energy of the photon flux of the single spectra of the GRB \citep[for the Swift-BAT case, see e.g. Fig. 7 of][]{Band2003}, we find that at $z=8.2$ the observed peak energies of any spectrum will be lowered by a factor $(1+0.54)/(1+8.2)$. Our procedure takes this further effect of to the cosmological redshift into account also.

\begin{figure}
\centering
\includegraphics[width=\hsize,clip]{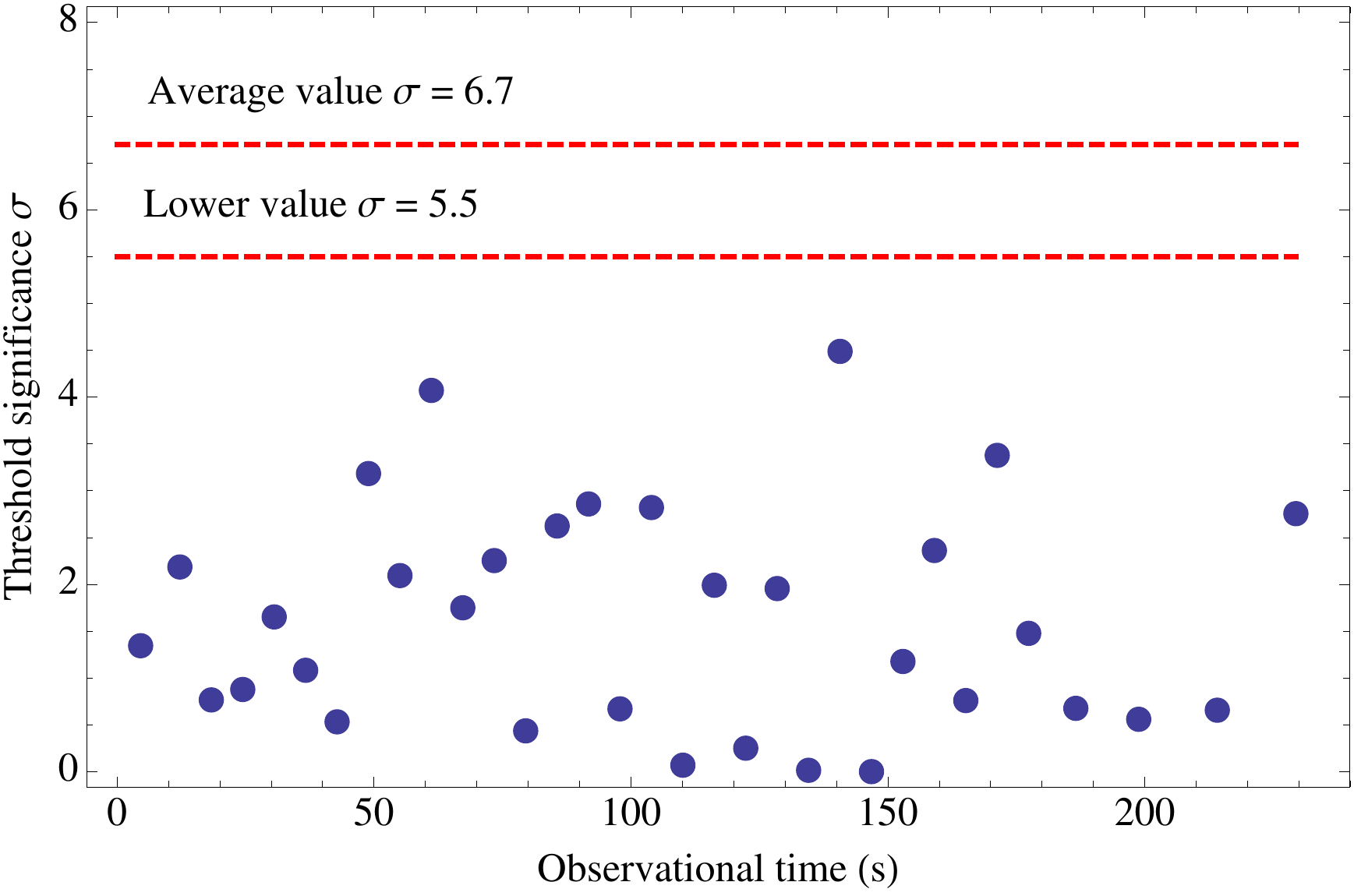}
\caption{Threshold significance $\sigma_0$ computed using the treatment of \citet{Band2003} for any single time-resolved spectra of the first emission episode in GRB 090618, as if they were emitted at redshift 8.2. The dashed lines correspond to the values for the threshold significance of $\sigma_0 = 5.5$ and $\sigma_0 = 6.7$ . }
\label{fig:no1}
\end{figure}

Since the threshold significance of Swift BAT is variable from a minimum value of $\sigma_0$ = 5.5 up to a maximum value of 11\footnote{http://swift.gsfc.nasa.gov/about\_swift/bat\_desc.html}, with an average value of $\sigma_0$ = 6.7, the results of this first analysis suggest that an Episode 1 similar to the one of GRB 090618 would not have been detected in GRB 090423 (see Fig. \ref{fig:no1}).

\begin{table*}
\centering
\begin{tabular}{lcccccc}
\hline\hline
  & $\alpha$ & $\beta$ & $E_{p,i}$ & $norm.$ & $\tilde{\chi}^2$ & $\Delta t_{obs}$ \\
  & & $(keV)$ & $(keV)$ & $(ph/cm^2/s/keV)$ & & $(s)$ \\
 \hline
 090618 & -0.66 $\pm$ 0.57 & -1.99 $\pm$ 0.05 & $284.57 \pm 172.10$ & 0.3566 $\pm$ 0.16 & 0.924 & 6.1 \\
 090423 & -0.78 $\pm$ 0.34 & -3.5 $\pm$ 0.5 & $433.6 \pm 133.5$ & 0.015 $\pm$ 0.010 & 0.856 & 10.4\\
 \hline
 \end{tabular}
\caption{Results of the spectral fits of the T90 duration of GRB 090423 and of the $\Delta t_{A,obs}$ time interval for GRB 090618. The latter is computed in a time interval corresponding to the one expected to be observed if GRB 090618 is transposed at the redshift $z$ = 8.2, and in the observed energy range (89.6 - 896) keV.}\label{tab:no1}
\end{table*}

\section{Detection of Episode 2 and its analysis}

\begin{figure}
\includegraphics[width=0.95\hsize,clip]{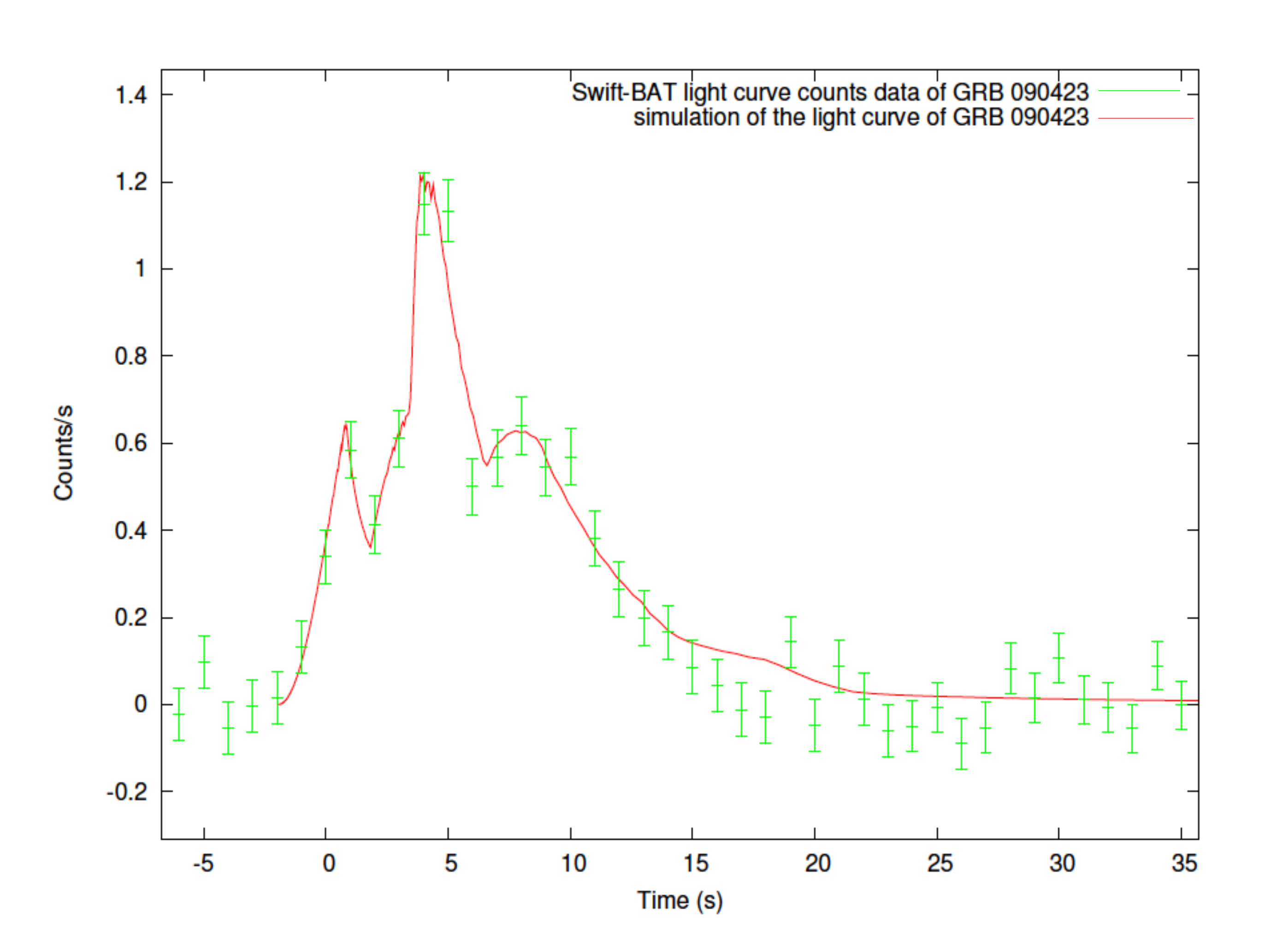}
\caption{Swift-BAT (15-150 keV) light curve emission of GRB 090423. The red line corresponds to the simulation of the GRB emission in the fireshell scenario \citep{Izzo2010}.}
\label{fig:no2c}
\end{figure}

Episode 2 emission of GRB 090423, detected by \textit{Swift}-BAT, was examined in the context of the fireshell scenario \citep{Izzo2010,Ruffini2011}. A Lorentz Gamma factor of $\Gamma \sim 1100$ and a baryon load $B = 8 \times 10^{-4}$ were obtained. The simulations of the observed spikes in the observed time interval (0 - 440) s lead to homogeneous circumburst medium ($\mathcal{R} = 10^{-8}$, see \citealp{Bianco2005a} for a complete description), and an average density of $10^{-1}$ particles cm$^{-3}$.
The simulation of the GRB 090423 emission is shown in Fig. \ref{fig:no2c}.

\begin{figure}
\includegraphics[width=0.93\hsize,clip]{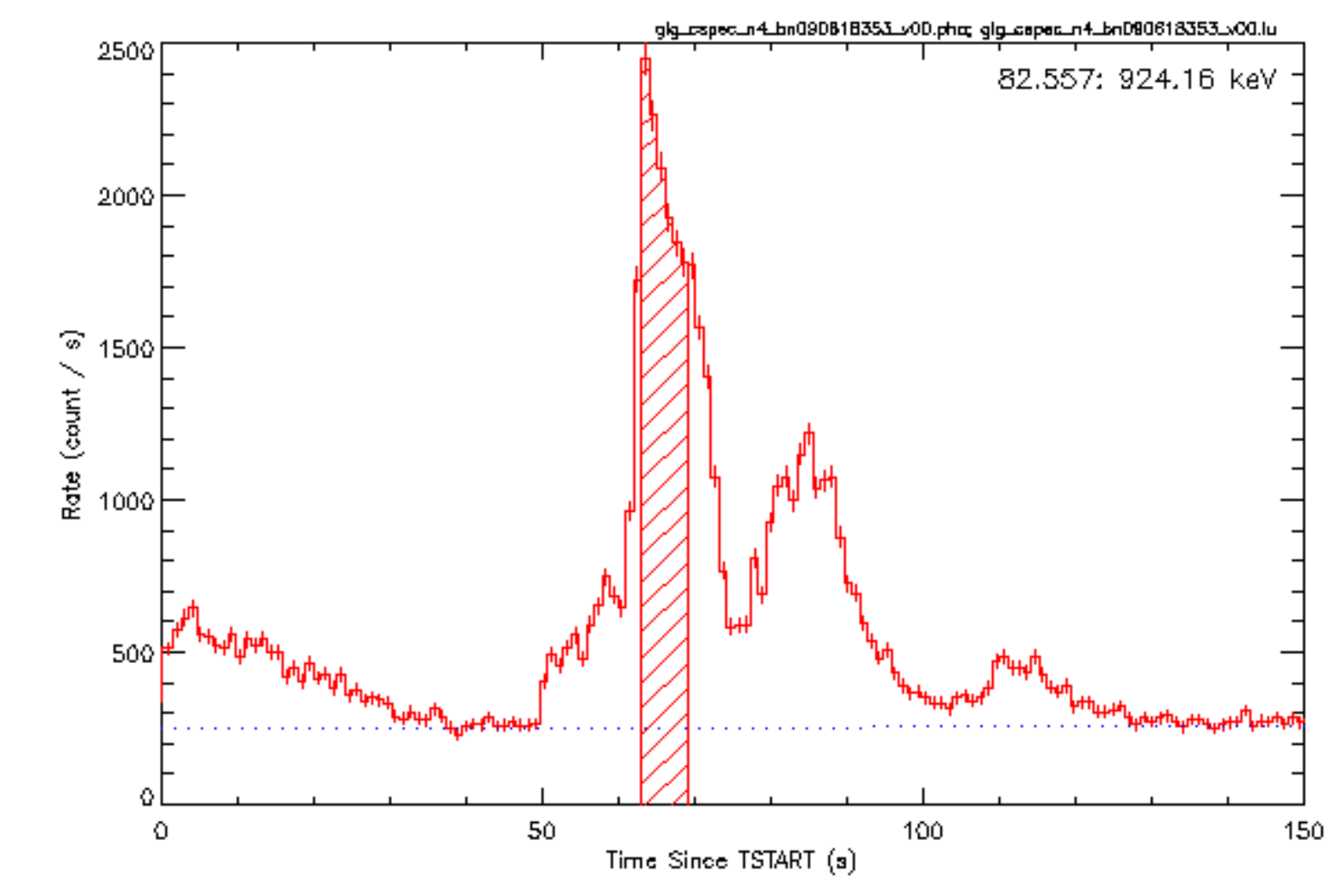}
\caption{Light curve of Episode 2  in GRB 090618, ranging from 50 to 150 s. The dashed region represents the portion which would have triggered the Swift BAT if this GRB had been at the redshift $z$ = 8.2. The observed duration of that interval is approximately $\Delta t \simeq 6$ s. The results obtained in Fig. \ref{fig:no2c}, when scaled to $z=0.54$, provide $\Delta T \simeq 3$ s.}
\label{fig:no2}
\end{figure}

\begin{figure}
\includegraphics[width=0.95\hsize,clip]{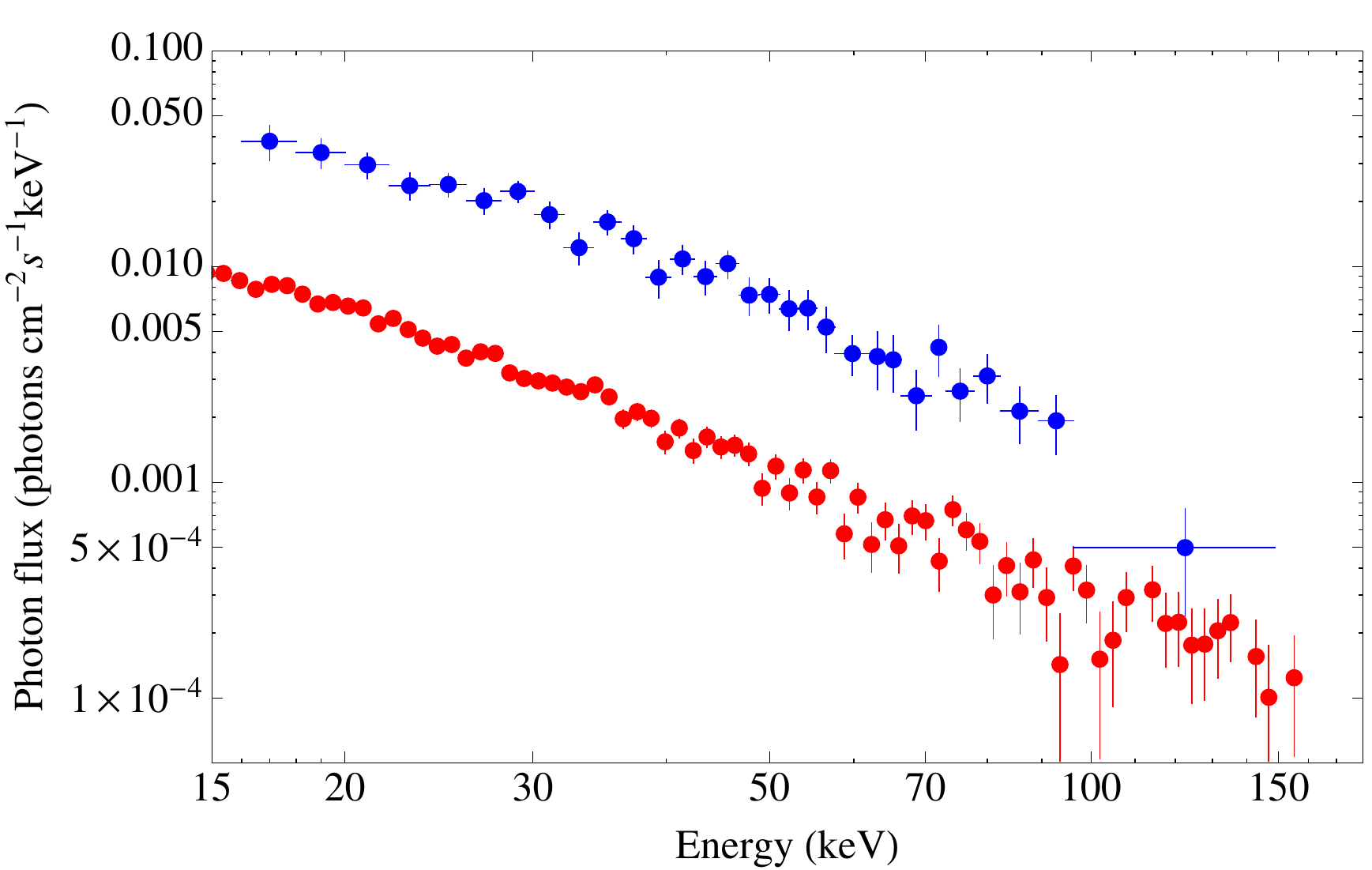}
\caption{Spectra of GRB 090423 (blue data) and of the spectrum of the emission of GRB 090618 (red data) considered as possible Episode 2 if GRB 090618 had been observed at $z$ = 8.2. The low-energy photon index is $\approx -0.8$, which corresponds to the expectations from the Fireshell scenario \citep{Ruffini2011,Patricelli2011}.}
\label{fig:no2b}
\end{figure}

We can now compare and contrast the emission observed in GRB 090423, expressed at $z = 8.2$ \citep[see Fig. \ref{fig:no2c},][]{Izzo2010}, and the portion of the emission of GRB 090618 if observed at $z = 8.2$, \citep[see Fig. \ref{fig:no2},][]{Izzo2012b}. In view of the Swift-BAT threshold, only the dashed region in Fig. \ref{fig:no2b}, lasting 6 s, would be detectable. The observed flux in Fig. \ref{fig:no2c} and the one of the dashed region in Fig. \ref{fig:no2b} will be similar when compared in a common frame.

For the above considerations, the analysis presented in the previous section can be applied to Episode 2 of GRB 090618. Assuming a detector threshold for Swift-BAT of $\sigma_0 = 6.7$, see Eq. \ref{eq:no2}, only the dashed region in Fig. \ref{fig:no2} is detectable when transposing GRB 090618 at $z=8.2$. In the observer frame, this emission corresponds to the time interval ($T_{0,G} + 63.0$, $T_{0,G} + 69.1$) s, with $T_{0,G}$ the trigger time of Fermi GBM data of GRB 090618. This time interval, at $z$ = 8.2, has a duration $\Delta t_{A,obs}$ = $\Delta t_{obs} \times 5.97$ = 36.4 s, owing to the time dilation by the cosmological redshift $z$ (see Fig. \ref{fig:no2c}). The remaining emission of GRB 090618 is unobservable, since below the threshold of the Swift-BAT detector. We note that $\Delta t_{A,obs}$ is quite comparable to the observed duration of GRB 090423 (see Fig. \ref{fig:no2c}). 

We turn now to comparing and contrasting the spectral energy distributions in the rest frame of the two GRBs. We consider the spectrum of GRB 090618 in the energy range (89.6 - 896) keV, which corresponds to the Swift BAT band (15 - 150) keV in the rest frame of GRB 090423. As for the time interval in GRB 090423, we consider the observational time interval (63.0 - 69.1) s, determined from applying Eq. \ref{eq:no2} to the entire Episode 2 of GRB 090618 (see the dashed region in Fig. \ref{fig:no2}).  We fitted the spectral emission observed in GRB 090423 with a Band function \citep{Band1993}, and the results provide an intrinsic peak energy $E_{p,i}  = (284.57 \pm 172.10)$ keV (see Table \ref{tab:no1}).  The same model provides for the spectral emission of GRB 090423, in the $T_{90}$ time duration, an intrinsic peak energy of $E_{p,i} = (433.6 \pm 133.5)$ keV. However, the break in GRB 090423 is steeper, while in GRB 090618 it is more shallow. This is  clear in Fig. \ref{fig:no2b}, where we show the spectra of both GRBs that are transformed to a common frame, which is the one at redshift $z=8.2$. Very likely, the difference in the steepening at high energies is related to the structure of the circumburst medium (CBM): the more fragmented the CBM, the larger the cutoff energy of the fireshell spectrum \citep{Bianco2005a}. Another important result is that the low energy index $\alpha$ is quite similar in both GRBs. This agrees with the expectation from the fireshell scenario, where a photon index of $\approx$ -0.8 is expected in the early emission of a GRB  \citep{Patricelli2011}.

The isotropic energy emitted in the time interval delineated by the dashed region in Fig. \ref{fig:no2} has been computed to be $E_{iso} = 3.49 \times 10^{52}$ erg, which is very similar to the one computed for the $T_{90}$ duration, in the same energy range, for GRB 090423, $E_{iso} = 4.99 \times 10^{52}$ erg.  

\section{Striking observations of Episode 3}

That in long GRBs the X-ray emission, observed by \textit{Swift}-XRT in energy range $0.3$--$10$ keV, presents a typical structure composed of a steep decay, a plateau phase and a late power-law decay, was clearly expressed by Nousek, Zhang and their collaborators \citep{Nousek2006,Zhang2006}.
This structure acquires a special meaning when examined in the most energetic sources, $E_{iso} = 10^{52} - 10^{54}$ erg, and leads to the fundamental proof that GRB 090423 is a BdHN source.

\begin{figure*}
\centering
\includegraphics[width=0.90\hsize,clip]{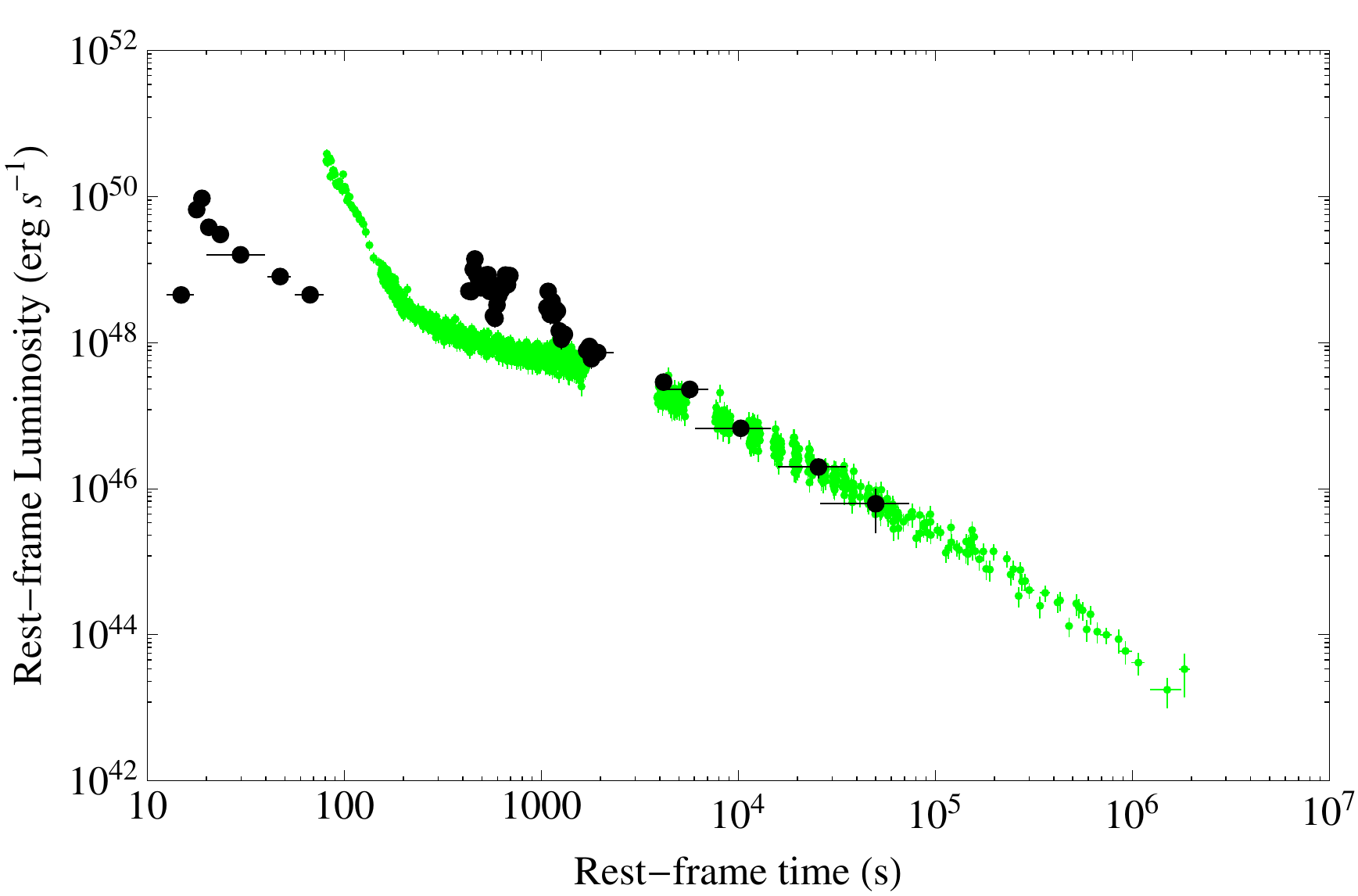}
\caption{Behavior of the Episode 3 luminosity of GRB 090423 (black dots) compared with the prototype case of GRB 090618 (green data).}
\label{fig:no3}
\end{figure*}

It has only been after applying the IGC paradigm to the most energetic long GRBs associated to SNe that we noticed the most unique characterizing property of the BdHN sources: while the steep decay and the plateau phase can be very different from source to source, the late X-ray power-law component overlaps, when computed in the cosmological rest-frame (see \citealp{Pisani2013} and Fig. \ref{fig:no0}). This has become the crucial criterion for asserting membership of a GRB in the BdHN family. Indeed, when we report the late X-ray emission of Episode 3 in GRB 090423 at $z=8.2$, and GRB 090618 at $z=0.54$, we observe a complete overlapping at times longer than $10^4$ s, see Fig. \ref{fig:no3}.

\subsection{Recent progress in understanding the nature of Episode 3}

We recall: 
\begin{itemize}
\item[ a)] that the X-ray luminosity of Episode 3 in all BdHN sources presents precise scaling laws (see, e.g., Fig. \ref{fig:no0});
\item[ b)] that the very high energy emission all the way, up to 100 GeV, in GRB 130427A, as well as the optical one, follows a power-law behavior similar to the one in the X-ray emission described above. The corresponding spectral energy distribution  is also described by a power-law function \citep{Kouveliotou2013,RuffiniWang2014}. These results clearly require a common origin for this emission process in Episode 3; 
\item[ c)] that an X-ray thermal component has been observed in the early phases of Episode 3 of GRB 060202, 060218, 060418, 060729, 061007, 061121, 081007, 090424,100316D, 100418A, 100621A, 101219B, and 120422A \citep{Page2011,Starling2012,Friis2013}. In particular, this feature has been clearly observed in GRB 090618 and GRB 130427A \citep{RuffiniWang2014}. This implies an emission region size of $10^{12-13}$ cm in these early phases of Episode 3, with an expansion velocity of $0.1 < v/c < 0.9$, with a bulk Lorentz $\Gamma$ factor $\lesssim 2$  \citep{RuffiniMuccino2014}. 
\end{itemize}

The simultaneous occurrence of these three features imposes very stringent constraints on any possible theoretical models. In particular, the traditional synchrotron ultra-relativistic scenario of the Collapsar jet model \citep{Woosley1993,MeszarosRees2000} does not appear suitable for explaining these observational facts.

In \citet{RuffiniMuccino2014}, we have recently pointed out the possibility of using the nuclear decay of ultra-heavy nuclei originally produced in the close binary phase of Episode 1 by $r$-process as an energy source of Episode 3. There is the remarkable coincidence that this set of processes leads to the value of the power-law emission with decay index $\alpha$, similar to the one observed and reported in \citet{Metzger2010}. The total energy emitted in the decay of these ultra-heavy elements agrees with the observations in Episode 3 of BdHN sources \citep{RuffiniMuccino2014}. An additional possibility of process-generating a scale-invariant power law in the luminosity evolution and spectrum are the ones expected
from type-I and type-II Fermi acceleration mechanisms \citep{Fermi1949}. The application of these acceleration mechanisms to the BdHN remnant has two clear advantages: 1) for us, to fulfill the above-mentioned power laws, both for the luminosity and the spectrum; and 2) for Fermi, to solve the longstanding problem, formulated by Fermi in his classic paper, of identifying the injection source to make his acceleration mechanism operational on an astrophysical level.

\section{Conclusions}

The ansatz that GRB 090423 is the transposed of GRB 090618 at $z=8.2$ has passed scrutiny. It is viable with respect to Episodes 1 and 4 and has obtained important positive results from the analysis of Episodes 2 and 3:

\begin{itemize}
\item Episodes 1 and 4 have not been detected in GRB 090423. This is consistent with the fact that the flux of Episodes 1 and 4 of GRB 090618 should not be observed by the Swift-BAT detector or by the optical telescopes, owing to the very high redshift of the source and the current sensitivities of X-ray and optical detectors;
\item Episode 2 of GRB 090423 has definitely been observed by Swift-BAT: its observed emission is comparable 1) to energy emitted ($3.49 \times 10^{52}$ erg for GRB 090618 and $4.99 \times 10^{52}$ erg for GRB 090423), 2) to the observed time duration (34 s for the observable part of GRB 090618 when transposed to $z=8.2$ and 19 s for GRB 090423), and 3) to the spectral energy distribution: the low energy part of the spectra of both GRBs is consistent with the expectation of the fireshell model \citep{Patricelli2011}. There is a significant difference only in the high energy part of the spectrum of GRB 090423, where a cutoff is observed at lower energy than the one in GRB 090618. This can be explained, in the fireshell scenario, by the existence of a dense and homogeneous CBM \citep{Bianco2005a}, which is expected for bursts at high redshifts;
\item Episode 3 shows the striking feature of the overlapping of the late X-ray luminosities of Episode 3 in GRB 090618 and GRB 090423, when compared in their cosmological rest frames (see Fig. \ref{fig:no3}). This result confirms the extension of the relation presented in \citet{Pisani2013} for $z \leq 1$, all the way up to $z=8.2$.
 \end{itemize}

From an astrophysical point of view, all the above results clearly indicate that 
\begin{itemize}
\item[ a)] GRB 090423 is fully consistent with being a member of the BdHN family, and the associated SN did occur already at $z=8.2$: the possibility of having an evolved binary system about 650 Myr after the Big Bang is not surprising, since the lifetime of massive stars with a mass up to 30 M$_\odot$ is $\sim$ 10 Myr \citep{Woosley2002}, which is similar to expectations from normal Population II binary stars also at $z=8.2$, as pointed out by \citet{Belczynski2010}; 
\item[ b)] the FeCO core and the NS companion occurring at $z=8.2$ also implies the existence, as the progenitor, of a massive binary  $\sim 40-60$ M$_\odot$\footnote{http://nsm.utdallas.edu/texas2013/proceedings/3/1/Ruffini.pdf}. Such massive binaries have recently been identified in Eta Carinae \citep{Damineli2000}. The very rapid evolution of such very massive stars will lead first to a binary X-ray source, like Cen-X3 \citet[see, e.g., ][]{GurskyRuffini} and \citet{GiacconiRR}, which will further evolve in the FeCO with the binary NS companion. A similar evolution starting from a progenitor of two very massive stars was considered by \citet{Fryer1999} and by \citet{Bethe1998}, leading to the formation of binary NSs or postulating the occurrence of GRBs. They significantly differ from the IGC model and also differ in their final outcomes; 
\item[ c)] the results presented in this article open the way to considering the late X-ray power-law behavior in Episode 3 as a distance indicator and represents a significant step toward formulating a cosmological standard candle based on Episode 3 of these BdHN sources.
\end{itemize}

We turn now to fundamental issues in physics. 
\begin{itemize}
\item[ 1)] The traditional fireball jet model \citep{Meszaros2006} describes GRBs as a single phenomenon, originating in a collapsar \citep{Woosley1993} and characterized by jet emission moving at Lorentz $\Gamma$ factor in the range $\approx 200$ -- $2000$. This contrasts with the BdHN model where the GRB is actually composed of three different episodes that are conceptually very different among each other (see Fig. \ref{fig:Maxime}) : Episode 1 is non-relativistic, and Episode 2 is ultra-relativistic with Lorentz $\Gamma$ factor $\approx 200$ -- $ 2000$, Episode 3 is mildly relativistic, with $\Gamma \approx 2$.
\item[ 2)] The description of Episode 1, see Fig. \ref{fig:Jorge}, proposes the crucial role of the Bondi-Hoyle hypercritical accretion process of the SN ejecta onto the NS companion. This requires an urgent analysis of the neutrino emission pioneered in the classic papers of \citet{Zeldovich1972,Chevalier1993,Fryer1996}, and \citep{Fryer2009}.
\item[ 3)] The binary nature of the progenitors in the BdHN model and the presence of the specific scaling power laws in the luminosity in Episode 3 of GRB 090423, as well as in all the other sources of the ``golden sample'' (see Fig. \ref{fig:no0}) \citep{Pisani2013}, has led us to consider the decay of heavy nuclear material originating in $r$-processes \citep{RuffiniMuccino2014}, as well as  type-I and type-II Fermi acceleration mechanism as possible energy sources of the mildly relativistic Episode 3 \citep{RuffiniWang2014}.

\end{itemize}

\begin{acknowledgements}

We are very grateful to the referee for the very thoughtful considerations and advices which greatly improved the presentation of our manuscript. We thank David W. Arnett for fruitful discussions and support. We are also grateful to Elena Zaninoni and Gennaro De Tommaso for their comments on the manuscript. This work made use of data supplied by the UK \emph{Swift} Science Data Centre at the University of Leicester. GBP, ME, and MK are supported by the Erasmus Mundus Joint Doctorate Program by grant Nos. 2011-1640, 2012-1710, and 2013-1471, respectively, from the EACEA of the European Commission.

\end{acknowledgements}

\end{document}